# Plasmon enhanced second harmonic generation by periodic arrays of triangular nanoholes coupled to quantum emitters

Elena Drobnyh[1] and Maxim Sukharev[1,2]

[1]Department of Physics, Arizona State University, Tempe, Arizona 85287, USA

[2]College of Integrative Sciences and Arts, Arizona State University, Mesa, Arizona 85201, USA

**Abstract**: Optical properties of periodic arrays of nanoholes of a triangular shape with experimentally realizable parameters are examined in both linear and nonlinear regimes. Utilizing fully vectorial three-dimensional approach based on the nonlinear hydrodynamic Drude model describing metal coupled to Maxwell's equations and Bloch equations for molecular emitters we analyze linear transmission, reflection, and nonlinear power spectra. Rigorous numerical calculations demonstrating second and third harmonic generation by the triangular hole arrays are performed. It is shown that both the Coulomb interaction of conduction electrons and the convective term contribute on equal footing to the nonlinear response of metal. It is demonstrated that the energy conversion efficiency in the second harmonic process is the highest when the system is pumped at the localized surface plasmon resonance. When molecular emitters are placed on a surface of the hole array lineshapes of the second harmonic signal exhibits three peaks corresponding to second harmonics of the localized surface plasmon mode and upper and lower polaritonic states.

## 1. Introduction

Optical properties of nanostructures comprised of noble metals such as gold and silver have been a subject of intense research for many years due to various fascinating applications.[1-5] Such systems are also useful as tools to study fundamental properties of light and its interaction with matter well below the diffraction limit.[6] Although linear optics of plasmonic materials is still of great interest, the nonlinear optical phenomena greatly enhanced by surface plasmon resonances are fast emerging.[7-13]

Beyond metallo-dielectric nanoconstructs and their properties are exciton-plasmon materials involving molecular assemblies and plasmonic systems.[14, 15] Under certain conditions when the coupling strength between molecules and local electromagnetic field associated with the plasmon resonance exceeds all dissipation rates, such a system enters the strong coupling regime. It is under these conditions one cannot really distinguish between molecular emitters and the electromagnetic radiation crafted by metal, i.e. they become hybridized exhibiting properties of both matter and light.[16] Various research groups investigated how such systems respond to external intense probes assuming that molecules are the main source of the nonlinear response.[17-22] One may note, however, that combining the nonlinear response of metal with that of molecules in the strong coupling regime obviously requires treating both subsystems on equal footing.

When considering nonlinear response of metal nanostructures two models are usually employed. Either one includes phenomenological nonlinear susceptibilities with different values at metal/dielectric interfaces and inside metal[23, 24] or the nonlinear hydrodynamics Drude model.[25-31] The latter even though is somewhat limited as it cannot account for contributions of core electrons and phonons, leads to surprisingly good agreement with experiments.[26, 28] Moreover, such a model is not perturbative and takes into account sharp metal boundaries, which arguably are the main source of nonlinearities in plasmonic nanostructures.

In this paper we employ the semi-classical approach based on coupled Maxwell-Bloch equations to account for optical response of molecular emitters and couple their dynamics directly to the nonlinear hydrodynamic Drude model describing metal. To the best of our knowledge this is the first time when both molecules and periodic nanohole arrays are considered on equal footing in the nonlinear regime.

## 2. Physical model

The electromagnetic radiation is described classically in accordance with the Maxwell equations

$$\frac{\partial \mathbf{B}}{\partial t} = -\nabla \times \mathbf{E},$$
$$\varepsilon_0 \frac{\partial \mathbf{E}}{\partial t} = \frac{1}{\mu_0} \nabla \times \mathbf{B} - \frac{\partial \mathbf{P}}{\partial t},$$
(1)

where the dynamics of the current density (macroscopic polarization current), $\mathbf{J} = \frac{\partial \mathbf{P}}{\partial t}$, follows microscopic material models as defined below.

The optical response in spatial regions occupied by metal is considered using classical hydrodynamics describing conduction electrons.[29] The corresponding equations of motion include the second law of Newton for the electron velocity field, $\mathbf{v}_e$, with the convective derivative and the number density, $n_e$, and the continuity equation

$$m^*\left(\frac{\partial}{\partial t}+\mathbf{v}_e\cdot\nabla+\gamma\right)\mathbf{v}_e = e\mathbf{E}+e\mathbf{v}_e\times\mathbf{B},$$

$$\frac{\partial n_e}{\partial t}+\nabla\cdot(n_e\mathbf{v}_e)=0, \quad (2)$$

where $m^*$ is the effective mass of conduction electrons and $\gamma$ is the phenomenological decay constant.[26] Defining the current density as $\mathbf{J}=en_e\mathbf{v}_e$ we can formally integrate the continuity equation. Combining this with the first equation of motion in Eq. (2) one arrives at the following equation for the macroscopic polarization[28]

$$\ddot{\mathbf{P}}+\gamma\dot{\mathbf{P}}=\frac{n_0 e^2}{m^*}\mathbf{E}+\frac{e}{m^*}\dot{\mathbf{P}}\times\mathbf{B}-\frac{e}{m^*}\mathbf{E}(\nabla\cdot\mathbf{P})-\frac{1}{n_0 e}[(\nabla\cdot\dot{\mathbf{P}})\dot{\mathbf{P}}+(\dot{\mathbf{P}}\cdot\nabla)\dot{\mathbf{P}}]. \quad (3)$$

where $n_0$ is the equilibrium number density of conduction electrons. The last three terms on the right-hand side are the nonlinear contributions due to the magnetic Lorentz force, Coulomb interaction arising from the continuity equation, and the convective term. It is informative to introduce dimensionless variables in Eq. (3), $t'=\frac{ct}{p}$, $r'=\frac{r}{p}$, where $p$ is a characteristic length scale (such as incident wavelength or period, for instance). The resulting equation reads

$$\ddot{\mathbf{P}}_j+\frac{d}{c}\gamma\dot{\mathbf{P}}=\frac{n_0 e^2}{m^*}\frac{p^2}{c^2}\mathbf{E}+\frac{ep}{m^*c}\dot{\mathbf{P}}\times\mathbf{B}-\frac{ep}{m^*c^2}\mathbf{E}(\nabla\cdot\dot{\mathbf{P}})-\frac{1}{n_0 ep}[(\nabla\cdot\dot{\mathbf{P}})\dot{\mathbf{P}}+(\dot{\mathbf{P}}\cdot\nabla)\dot{\mathbf{P}}]. \quad (4)$$

It appears[32] that in bulk the leading nonlinear contributions are arising due to the Lorentz force and the convective term, while the last two terms in Eq. (4) are dominant at metal-dielectric interfaces and for thin metal films.[28] In order to account for all possible nonlinear contributions we numerically integrate Eq. (3) with no further assumptions. When discussing our main results related to the second harmonic generation, we explicitly check which nonlinear term dominates the dynamics by subsequently turning these terms on and off in our codes.

In the linear regime, Eq. (3) reduces to the well-known equation for the current density, $\mathbf{J}$, as usually obtained from the conventional Drude model[33]

$$\dot{\mathbf{J}}+\gamma\mathbf{J}=\varepsilon_0\Omega_p^2\mathbf{E}, \quad (5)$$

where the plasma frequency is defined as $\Omega_p = \sqrt{\dfrac{n_0 e^2}{\varepsilon_0 m^*}}$. In our calculations we use the following parameters corresponding to silver: $\Omega_p = 8.28$ eV, $\gamma = 0.048$ eV, and $m^* = 0.99 m_e$.

To study strong coupling between plasmons and molecular assemblies and its consequences on second harmonic generation we employ rate equations that govern the dynamics of the macroscopic polarization in spatial regions with molecules[15]

$$\frac{dn_{ex}}{dt} + \Gamma n_{ex} = \frac{1}{\hbar \omega_{eg}} \mathbf{E} \cdot \dot{\mathbf{P}}_m,$$

$$\ddot{\mathbf{P}}_m + (\Gamma + 2\gamma^*)\dot{\mathbf{P}}_m + \omega_{eg}^2 \mathbf{P}_m = -\frac{2\omega_{eg}\mu_{eg}^2}{3\hbar}(2n_{ex} - n_0)\mathbf{E}, \tag{6}$$

where $n_{ex}$ is the number density of molecules in the excited state, $n_0$ is the total number density of molecules, $\omega_{eg}$ is the molecular transition frequency, $\mu_{eg}$ is the transition dipole, $\Gamma$ is the radiationless decay rate of the excited state, and $\gamma^*$ is the pure dephasing rate. In our simulations we vary the transition frequency and the number density, while keeping the other parameters fixed: $\mu_{eg} = 10$ D, $\Gamma = 4.1 \times 10^{-3}$ eV (corresponding to 1 ps), and $\gamma^* = 5.9 \times 10^{-3}$ eV (700 fs).

The resulting set of coupled equations (1), (3), and (6) constitutes the basis for considering molecules and plasmonic materials on equal footing and combining their response to external excitation in both linear and nonlinear regimes. One can build on this model to account for ro-vibrational degrees of freedom of molecules,[34] for instance, or lasing by implementing three- or four-level molecular transition schemes.[18, 35] We note that the presented model fully (within the range of applicability of the classical hydrodynamics model) accounts for the nonlinear optical response of metal directly altered by molecules in the self-consistent manner. One can also improve the molecular part of the model (6) by including corresponding master equation that governs the dynamics of the density matrix thus taking into account coherences for each molecule. This approach however becomes very involved since it is necessary to include additional electronic states with different magnetic numbers.[36]

Maxwell's equations are discretized in space and time following conventional finite-difference time-domain (FDTD) method using staggered griding.[37] We compute components of the macroscopic polarization, **P**, at the same spatial locations as the corresponding components of the electric field, **E**. In the Yee computational cell defined by three indexes (i, j, k), the three components of both **E** and **P** are evaluated at the locations of (i + ½, j, k), (i, j + ½, k), and (i, j, k + ½), respectively. To simulate open boundaries, we implement convolution perfectly matched layers (CPML).[38] The external excitation is simulated using the total field/scattered field approach. The linear response is obtained via the short pulse method with the incident pulse duration of 0.36 fs allowing one to obtain linear reflection/transmission/absorption spectra in a single FDTD run.[39] The nonlinear power spectra are evaluated using a 100 fs long incident pulse with the total propagation time of 500 fs. The numerical convergence in both linear and nonlinear regimes is achieved for a spatial resolution of 1.5 nm and a time step of 0.0025 fs. Our home-build codes are parallelized following the three-dimensional domain decomposition method.[40] All simulations were performed at AFRL/ERD DSRC HPC centers. For convergence test runs we utilized our local multiprocessor cluster. A typical number of computing cores used for a single production run varied between 12×12×11 = 1584 to 12×12×16 = 2304. Executions times were between 10 minutes, when simulating linear spectra, and 75 to 85 minutes when computing nonlinear power spectra.

## 3. Results and discussion

We consider periodic arrays of triangular holes in a thin silver film as schematically depicted in the inset of Fig. 1a. Although one may investigate second harmonic generation from circular holes, plasmonics of which is well understood,[41, 42] the magnitude of the signal is noticeably lower compared to the one generated by non-symmetric holes.[10] 280 nm thick silver metal film is placed on semi-infinite glass substrate with refractive index of 1.5 and is covered by a 100 nm thick polyvinyl alcohol (PVA) layer with the same refractive index.[43] It is also assumed that the areas inside holes is not filled with any dielectric. In this paper we study square arrays with the periodicity of 346.5 nm. The triangular holes have a side of 220 nm and the base of 200 nm. The system is excited by a plane wave at normal incidence emanating from the air side. Our simulations show that the linear transmission/reflection exhibits complex set of modes with most of them being located toward the ultraviolet part of the spectrum. The detailed analysis of local electromagnetic fields associated with these resonances reveal a multipole character of high energy modes.[44] We thus concentrate our interpretation efforts on understanding the physical nature of the three lowest (in energy) modes since they have three distinct physical features (electromagnetic energy spatially localized inside metal, on a boundary between metal and PVA, and inside PVA). Such spatial localizations should affect harmonic generation processes differently as discussed below.

Fig. 1 summarizes our main results on linear optical response. First, we examine the influence of sharp corners of holes on linear absorption as shown in Fig. 1a. Here we vary the curvature, $R$, of three corners of the hole and calculate the absorption as a function of the incident photon energy. We note that smooth corners result in a clear blue shift of all resonant modes, which can be explained by the fact that the volume of the hole decreases with the increasing curvature (the area of the hole decreases with higher $R$). Another feature observed is the mode near 1.75 eV, which appears for sharp corners and is hardly noticeable for $R = 20$ nm. This mode fades away when the curvature increases indicating that this feature is clearly due to the lightning rod effect as seen in many other plasmonic systems, where the sharp edge effect leads to knee-type resonances.[45, 46] Additionally we observe the enhancement and blue-shift of the high energy mode seen near 2.7 eV. Below we elaborate on the physical properties of the observed modes, however it is worth emphasizing that the high energy mode at 2.7 eV is enhanced noticeably by holes with round corners. A tincture of this mode is also seen for the sharp corners case at 2.56 eV.

The linear transmission, reflection, and absorption for the array of holes with round corners are plotted in Fig. 1b for the incident field polarized along $x$-axis (see inset of Fig. 1a with the definition of our system of coordinates). Three distinct resonant modes are seen at 1.99 eV, 2.54 eV, and 2.71 eV (resonant frequencies extracted from corresponding absorption spectra). Two low frequency resonances appear in both transmission and reflection spectra indicating that these modes have plasmonic character. The lowest frequency mode at 1.99 eV has a Q-factor of 12, which is typical for a plasmon resonance. Two other modes located at 2.54 eV and 2.71 eV, respectively, are more complex and worth discussing further. First, we note that the mode at 2.71 eV has nearly 0 transmission. Corresponding Q-factor of this resonance extracted from the absorption spectrum is 83. The intermediate mode at 2.54 eV is associated with the substantial transmission, but interestingly the reflection reaches a very low value of $3\times10^{-3}$. The Q-factor of this mode is 69.

To elucidate the physical nature of the observed modes we examine spatial distributions of their corresponding electromagnetic intensities (see Supplemental Material). The in-plane intensity distributions exhibit complex spatial characters with the energy localized mainly in-between the holes at 2.54 eV and 2.71 eV. The electromagnetic intensity calculated at 1.99 eV is predominantly localized at the edges of the holes and inside the metal. The longitudinal dependence shown in Fig. 1c makes it easier to see the plasmonic character of two of the three resonances. Here we plot the steady-state intensities computed at resonant frequencies corresponding to maxima of the absorption (Fig. 1b) as functions of the longitudinal coordinate, $z$. Two other coordinates are fixed such that we examine intensities along the detection line going through the hole (assuming that the origin is in the center of the cell (see the inset of Fig. 1a) the actual coordinates used are $x = -55$ nm and $y = 0$ nm). The initial enhancement is observed for all three modes near the boundary between the PVA and the hole. They are exponentially decreasing when going through the hole to the substrate. Another enhancement is seen on the output side near the substrate. However, this happens only for the 1.99 eV mode and the intermediate mode (2.54 eV).

The well-known expression for the resonant wavelength of the Bragg plasmon supported by a square array of holes at normal incidence,[47]

$$\lambda_{SPP} = \frac{\text{period}}{\sqrt{i^2 + j^2}} \sqrt{\frac{\varepsilon_{dielectric}\varepsilon_{metal}}{\varepsilon_{dielectric} + \varepsilon_{metal}}}, \qquad (7)$$

may be used to give a rough estimate for the lowest diffraction order $(i,j) = (1,0)$ resonant frequency. For the interface with glass/PVA we obtain 2.1 eV. We note, however that (7) does not take into account the shape of holes nor it accounts for a size of holes. Since the cross section of the holes has a triangular shape, we anticipate that Bragg plasmon modes must depend on in-plane incident polarization. To explore this dependence, we perform simulations varying the incident polarization and track corresponding resonances as shown in Fig. 1d.

Firstly, we note that the lowest frequency mode at 1.99 eV is merely affected exhibiting small variations of the shape of the resonance and its frequency, which shifts to lower frequencies when the polarization changes from $x$ to $y$ (we opted not to show this part of the spectrum as otherwise Fig. 1d would be very difficult to read). The main reason why the resonant frequency of the mode in question varies is the obvious non-symmetric shape of holes. We repeated simulations for holes having an equilateral triangular shape and observed that the lowest frequency resonance was nearly independent from the in-plane polarization. We thus conclude that the lowest frequency mode seen for $x$-polarization at 1.99 eV is a localized surface plasmon resonance (LSPR) associated with the individual hole. Two other modes are significantly dependent on the polarization as can be clearly seen from Fig. 1d. There are two modes exhibiting the surface plasmon character (showing noticeable transmission) when the incident polarization departs from *x*. Two plasmon resonances seen at 2.54 eV and 2.67 eV correspond to the first order Bragg plasmons. Their frequencies are different since the shape of the holes is not symmetric and the corresponding reciprocal lattice vectors along *x* and *y* are not the same.[48] We observe similar spectra with holes taken as equilateral triangles, i.e. the LPR mode is independent from the polarization while two Bragg modes are seen when the incident field is polarized in-between *x* and *y*. In summary, as seen in Fig. 1b the *x*-polarized incident field leads to excitation of the LPR mode (1.99 eV), the first order Bragg plasmon (2.54 eV), and the mode localized inside PVA with no surface plasmon character (2.71 eV), which may be interpreted as a waveguide mode in the PVA layer.

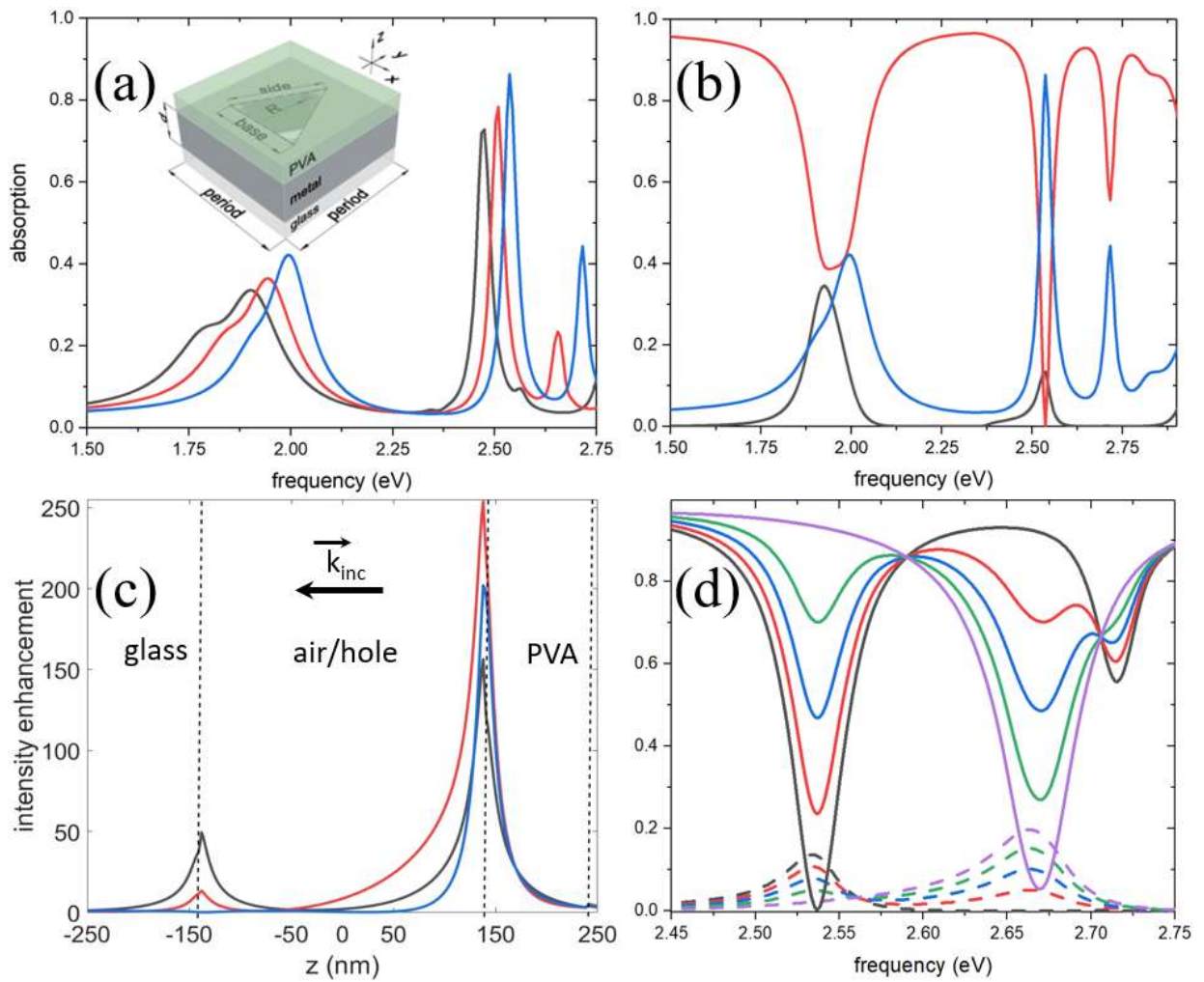

**Fig. 1.** Linear optics of triangle hole arrays. The inset in panel (a) shows the schematics of a unit cell of the array. The incident field propagates in the negative *z*-direction (from air) and is polarized in *xy*-plane. Panel (a) shows the linear absorption as a function of frequency for three values of the curvature *R* that defines how sharp the corners of the hole are. Black line is for $R = 1$ nm, red line is for $R = 10$ nm, and blue line is for $R = 20$ nm. The incident field is *x*-

polarized. Panel (b) presents linear transmission (black), reflection (red), and absorption (blue) as functions of frequency for the *x*-polarized incident field and $R = 20$ nm. Panel (c) shows the electromagnetic intensities calculated along *z*-axis that goes through the hole at three frequencies: (black) 1.99 eV, (red) 2.54 eV, and (blue) 2.71 eV. Other parameters are the same as in panel (b). Panel (d) shows linear transmission (dashed lines) and reflection (solid lines) as functions of frequency calculated at different incident field polarizations defined by the angle between the electric field and *x*-axis: (black) $0^0$ (i.e. *x*-polarization), (red) $30^0$, (blue) $45^0$, (green) $60^0$, (magenta) $90^0$ (i.e. *y*-polarization). All other parameters are the same as in panels (b) and (c).

Next, we examine how molecular emitters embedded in the PVA layer are coupled to either of the three resonant modes discussed above. Since the thickness of PVA is 100 nm and we consider molecules uniformly distributed inside PVA it is anticipated that the upper part of the layer may result in significant absorption without direct coupling to the corresponding electromagnetic mode. It is thus informative to examine absorption rather than transmission. Fig. 2 summarizes our results with molecules.

The optical absorption calculated for molecules resonant with the LSPR mode is shown in Fig. 2a. It is seen that that the absorption significantly increases at the resonant frequency with increasing molecular concentration. We observe the hybrid molecular-plasmon modes due to the strong coupling, which becomes obvious at the molecular concentration of $4\times10^{25}$ m$^{-3}$. The Rabi splitting (defined as the difference between resonant frequencies of the absorption for the hybrid exciton-plasmon modes) at this concentration is 78 meV. We also note that the absorption central peak is slightly blue shifted at high concentrations. See Supplemental Material for more details regarding molecules coupled to the LSRP mode. This interesting feature is more pronounced for molecules resonant with the Bragg plasmon (Fig. 2b).

The Rabi splitting due to the molecule-plasmon hybridization is significantly higher when the molecules are resonant at 2.54 eV (Bragg plasmon) and 2.71 eV (waveguide mode) as seen in Figs. 2b and 2c, respectively. In both cases, the Rabi splitting reaches 135 meV at the molecular concentration of $4\times10^{25}$ m$^{-3}$. Interestingly, in both cases the hybridization is very strong but the absorption peak at the corresponding molecular transition frequency is significantly pronounced for the waveguide mode (Fig. 2c), while the same peak is less affected by the molecular concentration in case of the Bragg mode. As noted above the absorption peak near corresponding molecular transition frequency shifts to higher frequencies at larger molecular concentrations, which means that the effective refractive index of the PVA layer with embedded molecules becomes smaller (rather than larger as one would expect). This is because at higher concentrations more molecules close to the metal/PVA interface are strongly coupled to the resonant electromagnetic mode resulting in less molecules absorbing at the molecular transition frequency. This effect is solely due to the strong coupling. If we are to tune the molecules away from a given resonant mode of the array the expected red shift of the absorption will be observed.

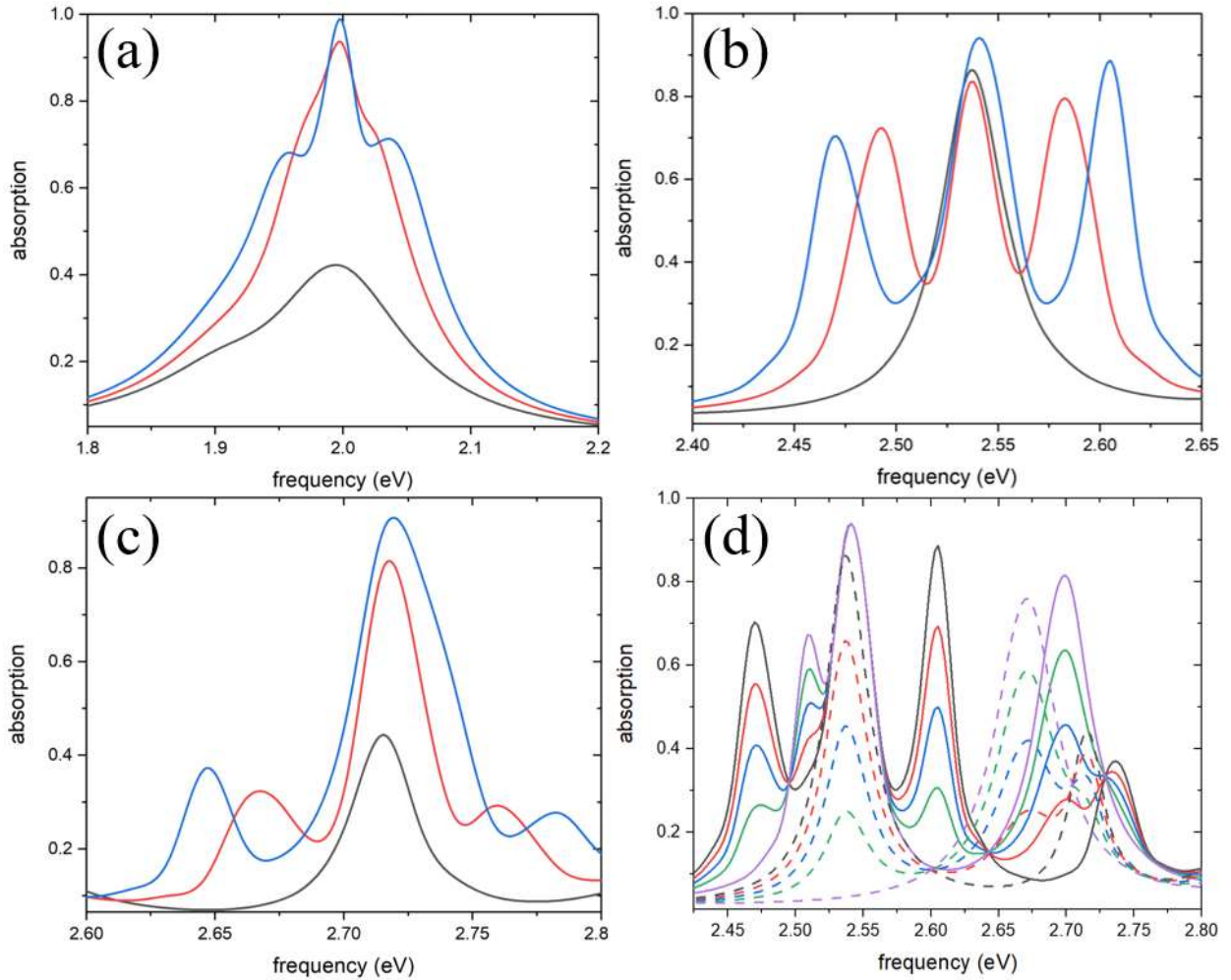

**Fig. 2.** Linear response of the exciton-plasmon arrays. This figure examines linear absorption of triangular hole arrays with two-level molecular emitters uniformly distributed inside the PVA layer. Panel (a) shows absorption for the array without emitters (black) and with emitters resonant at the LSPR frequency of 1.99 eV with the molecular concentration of $2\times10^{25}$ m$^{-3}$ (red) and $4\times10^{25}$ m$^{-3}$ (blue). Panel (b) shows absorption with molecules resonant at the first order Bragg plasmon frequency of 2.54 eV. The color scheme is the same as in panel (a). Panel (c) shows absorption for the molecules resonant at the frequency of 2.71 eV (the corresponding resonant mode is localized inside the PVA layer). The color scheme is the same is in panels (a) and (b). Panel (d) examines the polarization dependence of the absorption with the molecules resonant at 2.54 (Bragg mode). The dashed lines correspond to the array without molecules, the solid lines show the response of the array with molecules. Different incident field polarizations are defined by the angle between the electric field and *x*-axis: (black) $0^0$ (i.e. x-polarization), (red) $30^0$, (blue) $45^0$, (green) $60^0$, (magenta)

$90^0$ (i.e. *y*-polarization). The molecular density in panel (d) is $4\times10^{25}$ m$^{-3}$. The FWHM of the absorption for the stand-alone molecular layer is $4\times10^{-2}$ eV.

The polarization dependence of the absorption at the Bragg plasmon frequency is explored in Fig. 2d. Here we gradually change the in-plane polarization of the incident field keeping molecules resonant at 2.54 eV. Several interesting features are worth noting. First, compared to the bare array without molecules the central absorption peak near 2.54 eV is blue shifted for all polarizations. The effect we noted before in Fig. 2c. Secondly, the Rabi splitting does not depend on the polarization since the incident polarization does not affect the molecular concentration. The absorption for upper/lower polariton modes becomes smaller with the incident polarization rotating to *y* direction. This is the expected result since the first order Bragg plasmon along *x* has lower frequency compared to its counterpart excited at *y*-polarization (the effective distance between triangular holes along *x* is longer compared to that along *y* as we pointed out when discussing results in Fig. 1). Thirdly, we observe the absorption peak at 2.51 eV that becomes more pronounced as the incident polarization turns more toward *y*. For the intermediate polarization that excites both *x* and *y* Bragg plasmons one may see: (a) the polaritonic states associated with the strong coupling between molecules and *x*-Bragg plasmon; (b) the lower polariton state at 2.51 eV due to the coupling with the *y*-Bragg plasmon. Since the latter is far from the molecular resonance (2.54 eV) only a small portion of it is hybridized. The upper polariton part of this coupling is buried under the large absorption peak at 2.54 eV. We also note that the modes near 2.7 eV are shifted to higher frequencies compared to the case of the array without molecules as it was discussed above.

We now turn to examine how triangular hole arrays respond if subjected to an intense resonant laser pulse excitation. The first question to address is to see which nonlinear term on the left-hand side of Eq. (4) contributes most. Fig. 3a shows the normalized power spectrum calculated for the array pumped at the LSPR frequency of 1.99 eV. We first observe that turning on and off the magnetic part of the Lorentz force, $\dot{\mathbf{P}}\times\mathbf{B}$, has no feasible effect on the spectrum (the changes are seen only in the third significant figure and thus are not shown in the figure). Contributions from both the Coulomb term, $\mathbf{E}(\nabla\cdot\dot{\mathbf{P}})$, and the convective term, $[(\nabla\cdot\dot{\mathbf{P}})\dot{\mathbf{P}}+(\dot{\mathbf{P}}\cdot\nabla)\dot{\mathbf{P}}]$, have a noticeable impact on the signal for the second harmonic (calculations without the convective term underestimate the signal by the factor of 7) and the third harmonic (here the ratio of the signal evaluated using the full simulations and the one obtained from calculations without the convective term is 36). However, one may notice that neglecting either of these terms results in no dramatic changes nor in significant variations of the shape for each harmonic. Although the knee structure seen in the shape of the second harmonic is less pronounced when all the terms are considered, the change is small. If, however, both terms are dropped keeping only the Lorentz force resulting spectra (not shown) exhibit just a small tincture of the second harmonic only. We thus conclude that the main contributions to the nonlinear response come from terms with spatial derivatives.[28] One more comment is in order. Scalora et al.[28] developed a semi-analytical model based on Eq. (4). The nonlinear macroscopic polarization at the second harmonic was obtained (see Eq. (14) in [27]). It is seen that the polarization is governed by the magnetic part of the Lorentz force and the gradients that depend on the free electron susceptibility and local electric field at the fundamental frequency. As Fig. 2a demonstrates the gradients play the leading role in nanoscale systems and thus geometry becomes the major contributing factor.

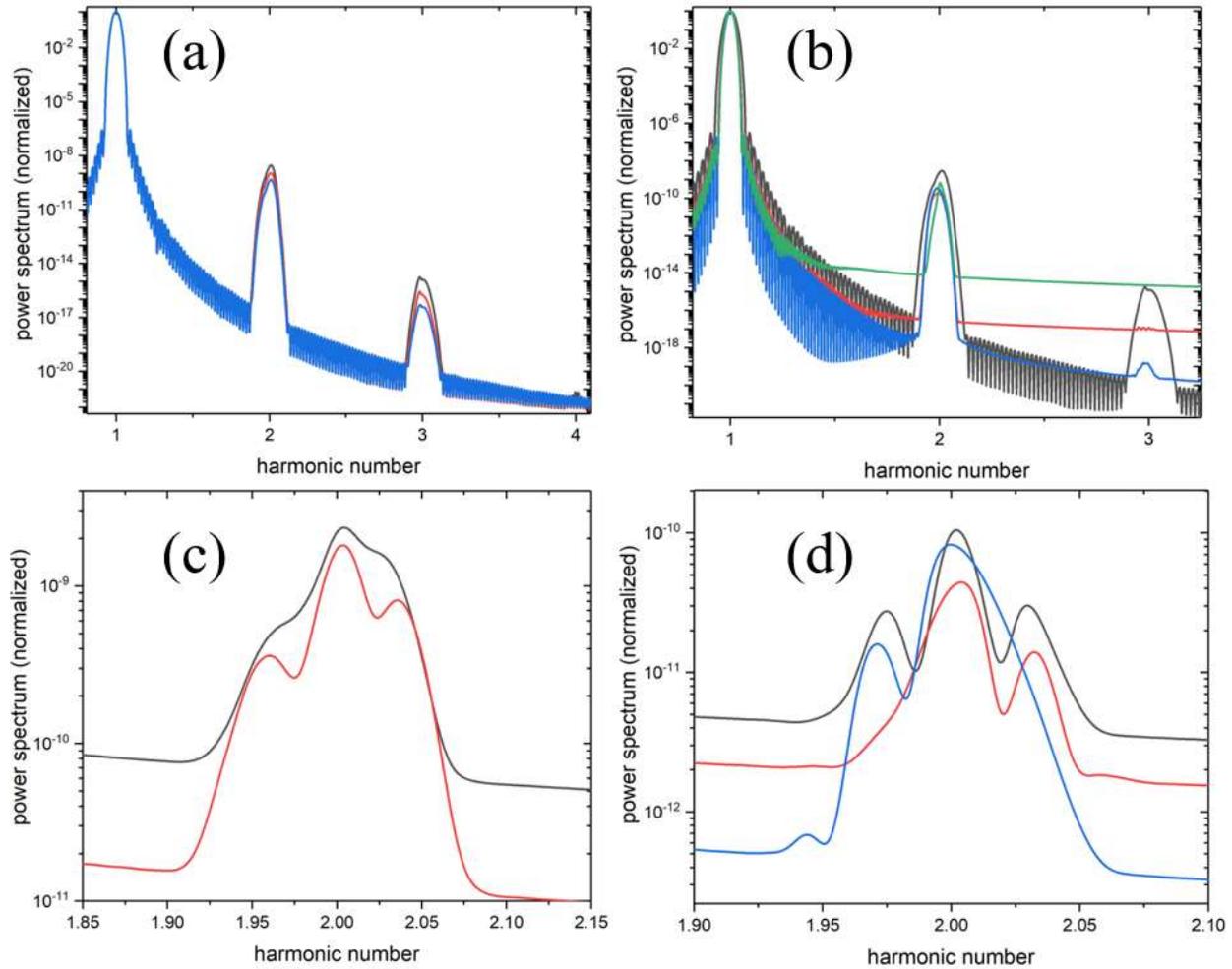

**Fig. 3.** The nonlinear response of the triangular hole arrays. Panel (a) shows the power spectra combing the transmitted and reflected energy due to the intense 100 fs laser pulse excitation of the array without molecules. Black line shows

complete model based on Eq. (3) with all terms included. Red line shows results of simulations when the Coulomb term (the third term in Eq. (3)) is dropped. Blue line shows results when the convective term (the last term in Eq. (3)) is neglected. The excitation pulse has the amplitude of $10^7$ V/m, is at the frequency of 1.99 eV, and is polarized along *x*. Panel (b) shows the normalized power spectra without molecules calculated for the pump frequencies of: (black) 1.99 eV, (red) 2.54 eV, (blue) 2.67 eV (here the pump is *y*-polarized), (green) 2.71 eV. The pump amplitude is the same as in panel (a). Panel (c) shows the power spectrum near the second harmonic of the pump. The spectrum is calculated for the array with molecules uniformly distributed inside the PVA layer. Black line shows results for the number density of molecules of $2\times10^{25}$ m$^{-3}$ and the red line is for the density of $4\times10^{25}$ m$^{-3}$. The molecular transition frequency is set at 1.99 eV. The incident pump is resonant at 1.99 eV, has the amplitude of $10^7$ V/m and is polarized along *x*. Panel (d) shows results of simulations similar to that in panel (c) but for three different pump frequencies: black line is for the pump at 2.54 eV (the first order Bragg plasmon), red line is for the pump at 2.49 eV (lower polariton), and blue line is for the pump at 2.58 eV (upper polariton). The pump amplitude and the polarization are the same as in panel (c). Molecular concentration is $2\times10^{25}$ m$^{-3}$. The molecular transition frequency is 2.54 eV.

Fig. 3b compares power spectra obtained by pumping the plasmonic array at four different resonant frequencies. The intensities of the second and third harmonics are significantly enhanced when the array is pumped at the LSPR frequency (1.99 eV). Pumping at either *x*- or *y*-Bragg plasmon resonances (2.54 eV and 2.67 eV, correspondingly) also leads to the second harmonic generation but the third harmonic is greatly suppressed. Interestingly the *y*-Bragg plasmon does show a clear signal at the third harmonic, while *x*-Bragg plasmon does not. When the array is pumped at the frequency of the guiding mode (2.71 eV) only the second harmonic is observed, which is not surprising since the resonant mode is mainly localized inside the PVA layer with a very small local field enhancement near metal interface. We also note that the lineshape of the second harmonic varies with the pump frequency. This is obviously due to the fact that the metal permittivity is frequency dependent. It is important to recall that our simulations are based on the nonlinear Drude model, which is valid only in a limited spectral range. To obtain a better qualitative description one may extend this model to a wider range of frequencies by phenomenologically including several Lorentz-type linear oscillators, which will provide a better fit for the linear part of the dielectric permittivity. While in the linear regime this is relatively straightforward,[37] one needs to be extra-careful in developing the analogy to Eq. (3) for the Drude-Lorentz model when entering the nonlinear regime. Research efforts in this direction are currently ongoing in our group.

Figs. 3c and 3d explore the second harmonic generation by triangular hole arrays with molecules uniformly distributed inside the PVA layer. We consider a simple two-level model for molecular emitters (Eq. (6)), which supports only odd harmonic generation due to the dipole selection rule. When molecules are not strongly coupled to any resonant modes of the array the even harmonic generation is solely due to metal. However, when the system enters the strong coupling regime there are significant changes to the lineshape of the second harmonic as seen in Fig. 3c. Here the system is pumped at the LSPR frequency with the molecules being resonant to that as well. A tincture of the second harmonic signals produced by the exciton-plasmon states at the concentration of $2\times10^{25}$ m$^{-3}$ is seen in Fig. 3c (black line). When the molecular concentration is increased three peaks are clearly visible. These correspond to the second harmonics of the lower polariton, the LSRP mode, and the upper polariton. One can calculate the Rabi splitting from this signal and compare it to the value of the Rabi splitting we extracted from Fig. 2a. The value calculated from Fig. 3c is 78 meV corresponds exactly to that obtained from linear absorption simulations. To understand how molecular transitions allowing only odd harmonic generation influence the second harmonic signal one may consider a simple analytical model we proposed elsewhere.[20] When an oscillator with a second order nonlinearity is strongly coupled to another linear oscillator the second order contribution to the displacement has a lineshape, which is significantly modified by the coupling strength of two oscillators.

We explore the influence of pumping at frequencies corresponding to hybrid states in Fig. 3d, when molecules are coupled to the Bragg mode at 2.54 eV. When pumped at 2.54 eV we again observe three peaks in the second harmonic signals. Furthermore, the Rabi splitting of 135 meV we calculated from this signal corresponds to that obtained from Fig. 2b. Two different scenarios are also considered. First, if pumped at the frequency of the lower polariton (Fig. 3d, red line) its second harmonic is noticeably enhanced and the second harmonic of the Bragg plasmon is seen as well. Secondly, when we drive the system at the upper polariton frequency in addition to its second harmonic and the corresponding plasmon peak we also see a small contribution from the lower polariton as well.

To explore angular dependence and the physical nature of the second harmonic generation enhanced by the LSPR mode we separately calculate horizontally and vertically polarized second harmonic signals. The measurement of those is performed in the far field zone along a circular detection contour (see Fig. 4a). Fig. 4 shows calculated angular diagrams. It is evident that the horizontally polarized second harmonic signal is left-right symmetric due to the mirror symmetry of the triangular shape. However, the vertical polarization exhibits highly enhanced directional radiation in the direction of the upper corner of the triangular hole indicating that the hole acts as a directional nonlinear antenna.

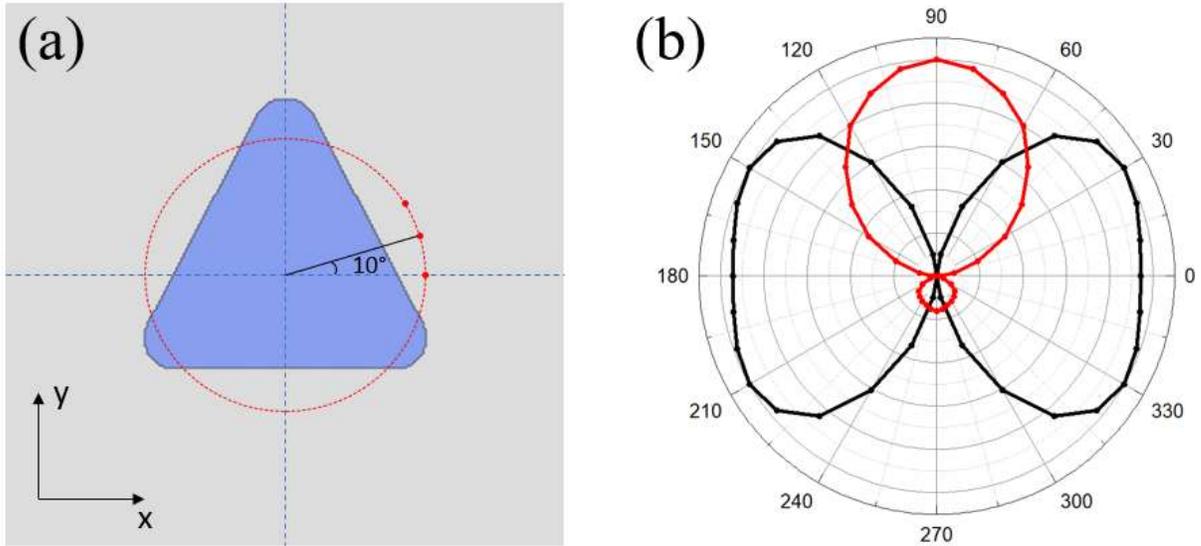

**Fig. 4.** Polarization dependence of the second harmonic signal. Panel (a) shows top view of the unit cells of the simulation domain. Red dashed circle with a radius of 100 nm is placed 588 nm above the PVA layer. It is comprised of 36 detection points equally spread along the circle with the angular displacement of 10°. Panel (b) shows horizontally, $|E_x(2\omega)|^2$, (black) and vertically, $|E_y(2\omega)|^2$, (red) polarized second harmonic signals for the triangular hole arrays without molecules detected along the detection contour shown in panel (a). The pump pulse has the amplitude of $10^7$ V/m with the duration of 100 fs, and is polarized along x. The pump frequency is 1.99 eV.

## 4. Conclusion

In summary, we investigated linear and nonlinear responses of periodic arrays of nanoholes of a triangular shape using experimentally realizable parameters. It was shown that such arrays exhibit a set of optical resonances. We discussed and pinpointed the physical nature of the first three modes. The modes in question were the localized surface plasmon mode, the first order Bragg plasmon mode, and the waveguided mode supported by a thin dielectric layer placed on the input side of the array. When resonantly coupled to molecular emitters periodic arrays show hybridization between molecular states and corresponding resonant plasmon modes. Significantly large Rabi splittings over 100 meV were found for molecules resonant with the first order Bragg plasmon modes. The latter also depend on the incident field polarization, which can be used as s tuning knob to modify the coupling strength between surface modes and molecules. We also observed an interesting effect, namely a blue shift of linear absorption peaks, which we assigned to the strong coupling between molecules and plasmons. We investigated how the arrays with and without molecules optically respond when driven by intense resonant laser pulses. Using the nonlinear hydrodynamic Drude model we showed that such systems support second and third harmonic generation. Local electromagnetic fields associated with the second harmonic exhibit complex spatial dependence. The electric field components induced by the metal are localized in-between triangular holes. If coupled to molecules under strong coupling conditions the second harmonic lineshapes are shown to be substantially modified exhibiting three peaks that correspond to the second harmonic signals at a driving frequency and upper and lower polaritons. This observation is supported by the simple analytical model, which we recently proposed elsewhere.[20]

## 5. Acknowledgements

The work is sponsored by the Air Force Office of Scientific Research under Grant No. FA9550-19-1-0009. The authors also acknowledge computational support through Department of Defense High Performance Computing Modernization Program. M.S. is grateful to the travel support provided by the Binational Science Foundation through Grant No. 2014113. The authors would like to thank Adi Salomon and Yehiam Prior for fruitful discussions.

# Supplementary Material:

# Plasmon enhanced second harmonic generation by periodic arrays of triangular nanoholes coupled to quantum emitters

Elena Drobnyh[1] and Maxim Sukharev[1,2]

[1]Department of Physics, Arizona State University, Tempe, Arizona 85287, USA

[2]College of Integrative Sciences and Arts, Arizona State University, Mesa, Arizona 85201, USA

1. **Avoided crossing for molecules coupled to the LSPR mode.**

Fig. S1 shows resonant frequencies of the upper and lower polaritons near the LSPR mode (1.99 eV). Calculations are performed by sweeping molecular transition frequency through the LSPR, corresponding resonant frequencies are extracted from the absorption spectra.

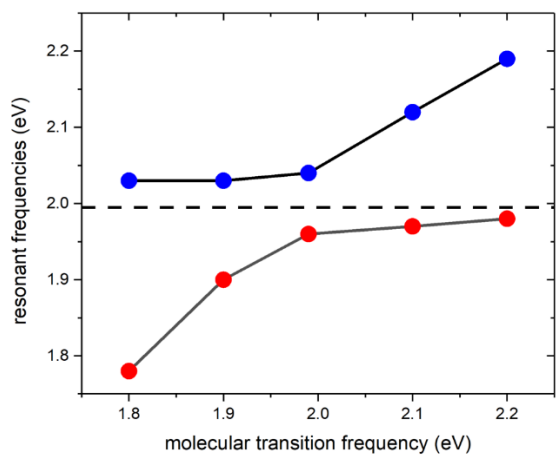

**Fig. S1**. Resonant frequencies of the lower (red circles) and upper (blue circles) polaritons as functions of the molecular transition frequency. The resonant frequencies are extracted from absorption spectra for the triangular hole array with molecules embedded in PVA. Horizontal dashed line shows the LSRP frequency of the bare triangular hole array. Molecular concentration is set at $4 \times 10^{25}$ m$^{-3}$.

2. **Local EM field spatial distributions for the LSR, the Bragg plasmon, and the guiding mode.**

Fig. S2 shows EM field components and the corresponding intensity as functions of X and Y distributions 20 nm above the input side of the array. Calculations are performed to obtain steady-state solutions of the Maxwell's equations when the array is driven by CW field at the LSPR frequency and is polarized along X. One can see that EM field is mostly localized near the upper edges of the hole. Spatial distributions corresponding to the Bragg plasmon mode are shown in Fig. S3. One can see that the field is localized in-between the holes as expected. Fig. S4 shows fields for the guiding mode.

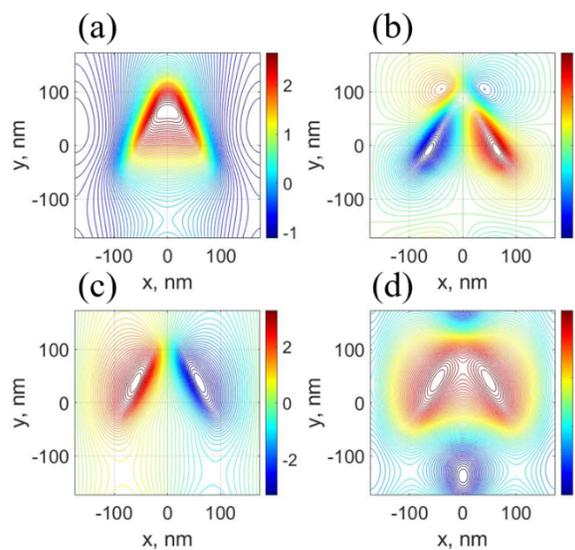

**Fig. S2.** Local electric field maps for the LSPR mode. The excitation pulse has the amplitude of 1.0 V/m, is at the frequency of 1.92 eV, and is polarized along x. Panels (a), (b), and (c) show electric field components ($E_x$, $E_y$, and $E_z$, respectively) as functions of x and y. Each electric field component is normalized with respect to the incident field amplitude. Panel (d) shows the electromagnetic intensity on logarithmic scale. Simulations are performed for a detection XY-plane placed inside the PVA 20 nm above the metal surface.

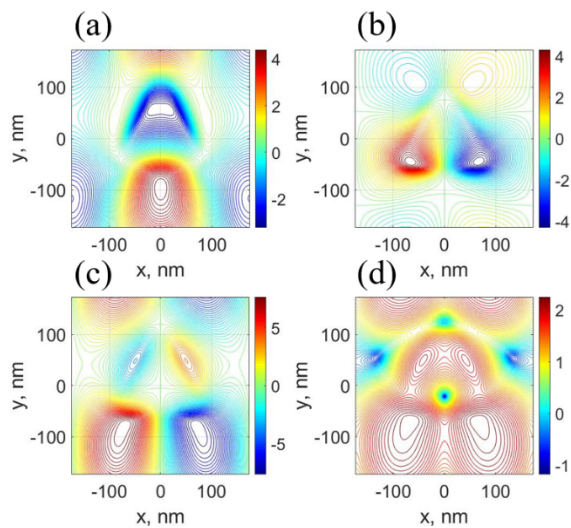

**Fig. S3.** Same as in Fig. S3 but for the Bragg plasmon mode, i.e. at 2.54 eV.

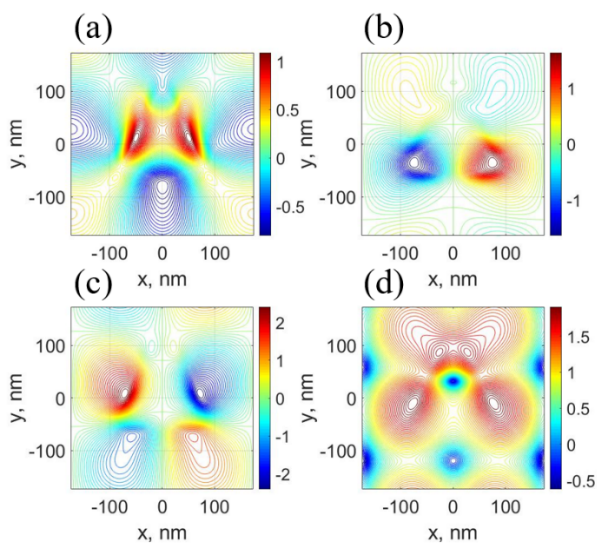

**Fig. S4.** Same as in Figs. S3 and S4 but for the guiding mode, i.e. at 2.72 eV.

## 3. Nonlinear fields in the triangular hole arrays

Having direct access to local electromagnetic field components we can examine spatial distributions of the second harmonic fields as well. Figs. S5, S6, S7 show steady-state distributions of the electromagnetic field and local intensities corresponding to the second harmonic of the LSPR mode (S5), the Bragg plasmon mode (S6), and the guiding mode (S7). It is interesting to note that the second harmonic of the LSPR is spatially spread across the entire unit cell, while other two cases show strong localization of the fields near the edges of the hole.

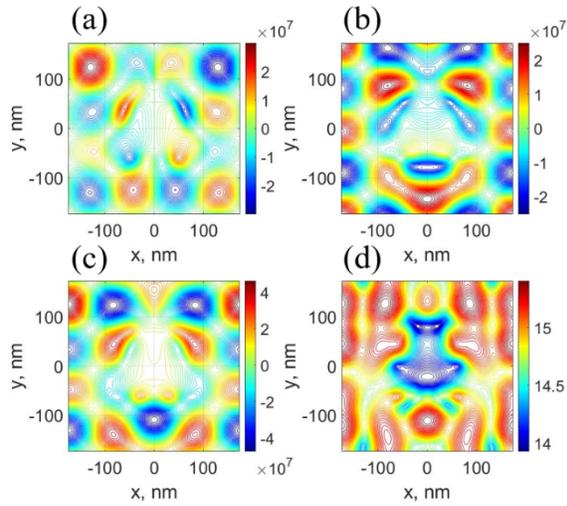

**Fig. S5.** Local field distribution calculated at the second harmonic of the pump at the LSPR frequency (1.99 eV) The pump has a peak amplitude of $10^7$ V/m and is horizontally polarized. Panels (a), (b), and (c) show electric field components ($E_x$, $E_y$, and $E_z$, respectively) as functions of x and y. Panel (d) shows the electromagnetic intensity on logarithmic scale. Simulations are performed inside the PVA layer at 20 nm above the metal surface.

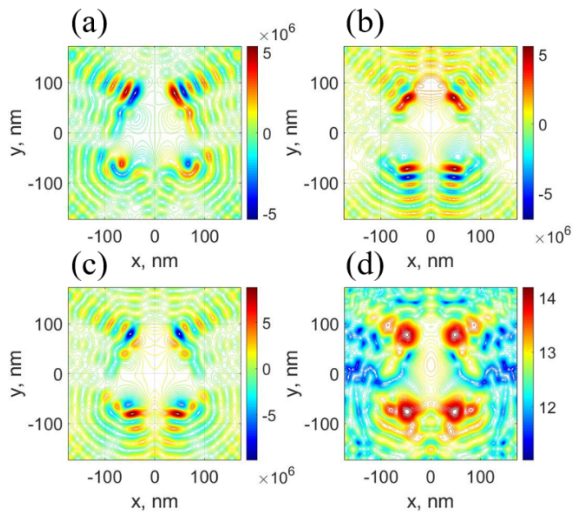

**Fig. S6.** Same as in Fig. S6 but for the Bragg plasmon at 2.54 eV.

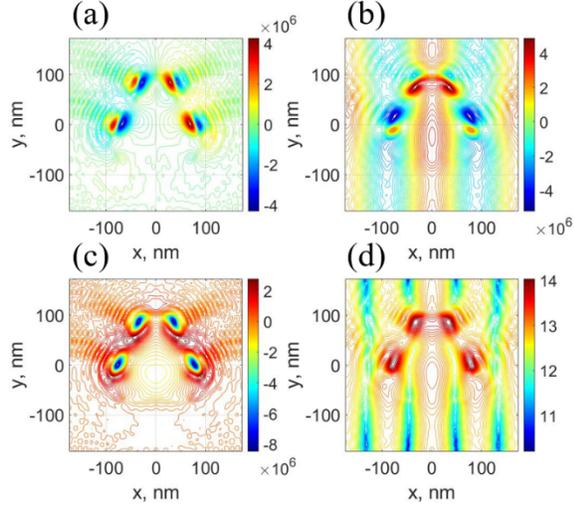

Fig. S7. Same as in Figs. S6 and S7 but for the guiding mode at 2.72 eV.

## 4. Angular properties of the second harmonic signal

Fig. S8 shows angular distributions of the horizontally and vertically polarized second harmonic signal for the Bragg plasmon mode and the guiding mode (see Fig. 4 in the paper for the LSPR distribution and details of calculations). The triangular hole arrays are pumped at 2.54 eV and 2.71 eV polarized along X (horizontal) and at 2.26 eV and 2.67 eV polarized along Y (vertical). Angular distributions indicate significant enhancement of the intensity at the lower corners of the triangular holes (Fig. S8a and S8d) and in the upper corner (Fig. S2b and S2c).

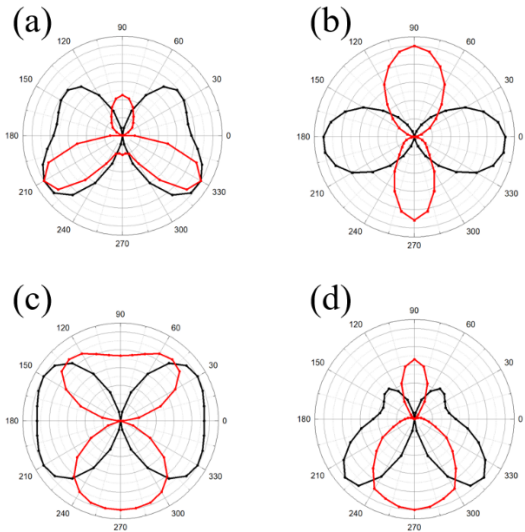

Fig. S8. Angular properties of the second harmonic signal: $E_x^2$ (black) and $E_y^2$ (red). There are 36 detection points, placed 588 nm above the PVA layer, lined up along the circle with the radius of 100 nm and the step is 10°. The pump amplitude is $10^7$ V/m with the duration of 100 fs. It is (a) polarized along X at 2.54 eV, (b) polarized along X at 2.71 eV, panel (c) polarized along Y at 2.26 eV, panel (d) polarized along Y at 2.67 eV.

## 5. Nonlinear fields in the circle hole arrays

Since our approach is based on direct integration of Maxwell's equations coupled to the equation on the macroscopic polarization, it accounts for a symmetry breaking in the direction of the incident field propagation. Thus we can in principle observe even harmonics at symmetric systems such as arrays of circular holes. Fig. S9 shows linear response of such an array. The setup is identical to the one used for the triangular hole arrays.

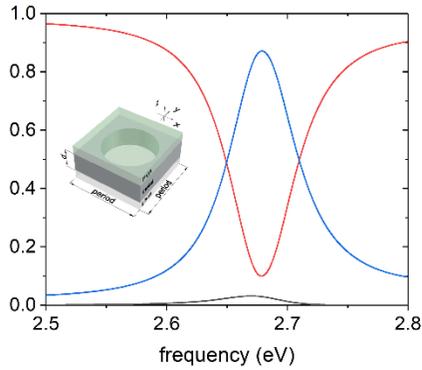

**Fig. S9.** The inset shows the schematics of a unit cell of the array. The incident field propagates in the negative $z$-direction (from air) and is polarized in $xy$-plane. The main figure shows linear transmission (black), reflection (red), and absorption (blue) as functions of the incident frequency. The radius of the circular hole is $R = 75$ nm.

The surface plasmon mode is observed at 2.67 eV. Fig. S10 shows corresponding LSR electromagnetic field distributions. When pumped at this frequency the array supports second harmonic fields, which are presented in Fig. S11.

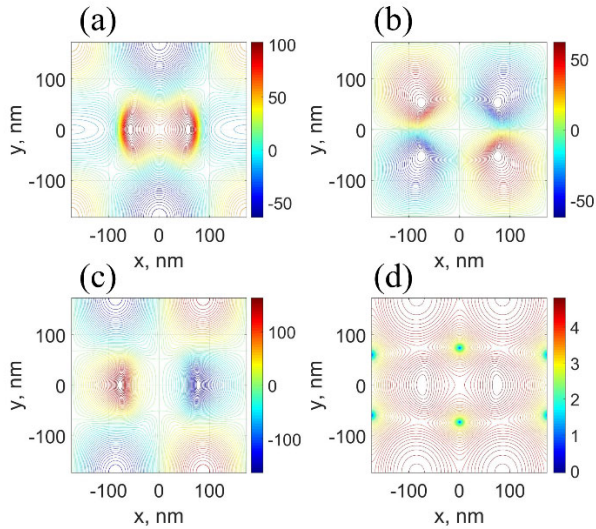

**Fig. S10.** Local electric field maps for the plasmon mode at 2.67 eV. The excitation pulse is polarized along x. Panels (a), (b), and (c) show electric field components ($E_x$, $E_y$, and $E_z$, respectively) as functions of x and y. Each electric field component is normalized with respect to the incident field amplitude. Panel (d) shows the electromagnetic intensity on logarithmic scale. Simulations are performed for a detection XY-plane placed inside the PVA 20 nm above the metal surface.

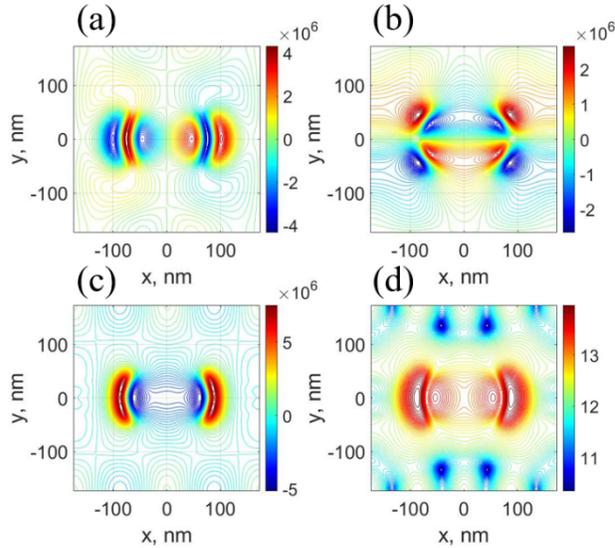

**Fig. S11.** Local field distribution calculated at the second harmonic of the pump centered at 2.67 eV. The pump has a peak amplitude of $10^7$ V/m and is horizontally polarized. Panels (a), (b), and (c) show electric field components ($E_x$, $E_y$, and $E_z$, respectively) as functions of x and y. Panel (d) shows the electromagnetic intensity on logarithmic scale. Simulations are performed inside the PVA layer at 20 nm above the metal surface.

Fig. S12 directly compares power spectra obtained for the triangular hole array (pumped at 1.99 eV) and the circular hole array considered above.

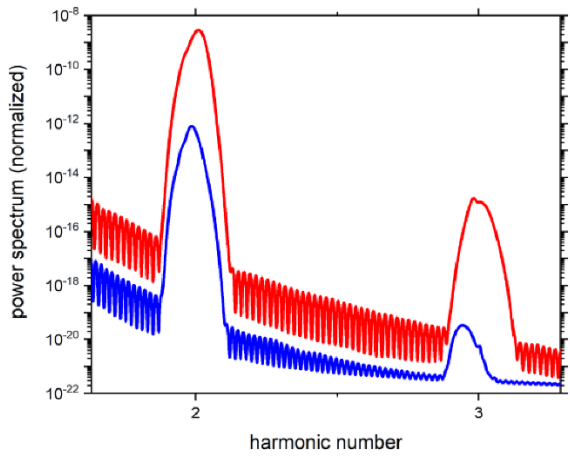

**Fig. S12.** Normalized power spectra calculated for the periodic arrays of circular (blue) and triangular (red) holes. The geometrical parameters for the arrays are the same as in Fig. 1 of our manuscript. Both arrays are pumped by a 100 fs laser pulse with the peak amplitude of $10^7$ V/m. The pump frequencies correspond to localized surface plasmon modes shown in Fig. 1 in this letter. The pump frequency for the circular hole array is at 2.67 eV corresponding the dipolar plasmon resonance seeing in Fig. S9.